\documentclass[conference,a4paper]{IEEEtran}
\IEEEoverridecommandlockouts

\usepackage[hidelinks]{hyperref}
\usepackage[cmex10]{amsmath}
\usepackage{amssymb,amsfonts}
\usepackage{dblfloatfix}

\usepackage[ruled,vlined]{algorithm2e}
\usepackage{graphicx}
\graphicspath{{Figures/PDF/}{Figures/PNG/}}

\usepackage{booktabs}
\usepackage{siunitx}
\usepackage[numbers,compress]{natbib}
\usepackage{texnames}
\usepackage{bm,bbm}
\usepackage{orcidlink}
\usepackage{multirow}
\usepackage{xcolor}
\definecolor{bestColor}{RGB}{255, 0, 0}    
\definecolor{secondBestColor}{RGB}{0, 0, 255} 
\definecolor{thirdBestColor}{RGB}{240,0, 240}  
\definecolor{qBestColor}{RGB}{0,215, 0}  

\definecolor{background_color}{RGB}{255,255,255}  

\definecolor{cnn_color}{RGB}{240,248,235}  
\definecolor{transformer_color}{RGB}{245,240,230}  
\definecolor{ours_color}{RGB}{255,225,240}  
\definecolor{input_color}{RGB}{255,240,240}      

\usepackage{colortbl}  

\newcommand{\best}[1]{\textcolor{bestColor}{\textbf{#1}}}      
\newcommand{\secondBest}[1]{\textcolor{secondBestColor}{\textbf{#1}}} 

\begin{document}
\title{HyFusion: Enhanced Reception Field Transformer for Hyperspectral Image Fusion
\thanks{\textit{(Corresponding author: Chih-Chung Hsu; Email: cchsu@gs.ncku.edu.tw)}}
}

\author{
    \IEEEauthorblockN{
        Chia-Ming Lee\orcidlink{0009-0004-6027-3083}\IEEEauthorrefmark{1},
        Yu-Fan Lin\orcidlink{0009-0000-9459-701X}\IEEEauthorrefmark{1}, 
        Yu-Hao Ho\orcidlink{0009-0001-0596-0380}\IEEEauthorrefmark{2},
        Li-Wei Kang\orcidlink{0000-0001-6529-3981}\IEEEauthorrefmark{2},
        Chih-Chung Hsu\orcidlink{0000-0002-2083-4438}\IEEEauthorrefmark{1},
    }
    \IEEEauthorblockA{
        \IEEEauthorrefmark{1}\textit{National Cheng Kung University}, Tainan, Taiwan (R.O.C.)\\
        \{zuw408421476@gmail.com, aas12as12as12tw@gmail.com, cchsu@gs.ncku.edu.tw\}
    }
    \IEEEauthorblockA{
        \IEEEauthorrefmark{2}\textit{National Taiwan Normal University}, Taipei, Taiwan (R.O.C.)\\
        \{i0981526423@gmail.com, lwkang@ntnu.edu.tw\}
    }
}

\maketitle
\begin{abstract}
Hyperspectral image (HSI) fusion addresses the challenge of reconstructing High-Resolution HSIs (HR-HSIs) from High-Resolution Multispectral images (HR-MSIs) and Low-Resolution HSIs (LR-HSIs), a critical task given the high costs and hardware limitations associated with acquiring high-quality HSIs. While existing methods leverage spatial and spectral relationships, they often suffer from limited receptive fields and insufficient feature utilization, leading to suboptimal performance. Furthermore, the scarcity of high-quality HSI data highlights the importance of efficient data utilization to maximize reconstruction quality. To address these issues, we propose HyFusion, a novel Dual-Coupled Network (DCN) framework designed to enhance cross-domain feature extraction and enable effective feature map reusing. The framework first processes HR-MSI and LR-HSI inputs through specialized subnetworks that mutually enhance each other during feature extraction, preserving complementary spatial and spectral details. At its core, HyFusion utilizes an Enhanced Reception Field Block (ERFB), which combines shifting-window attention and dense connections to expand the receptive field, effectively capturing long-range dependencies while minimizing information loss. Extensive experiments demonstrate that HyFusion achieves state-of-the-art performance in HR-MSI/LR-HSI fusion, significantly improving reconstruction quality while maintaining a compact model size and computational efficiency. By integrating enhanced receptive fields and feature map reusing into a coupled network architecture, HyFusion provides a practical and effective solution for HSI fusion in resource-constrained scenarios, setting a new benchmark in hyperspectral imaging. Our code will be publicly available.

\end{abstract}


\section{Introduction}\label{sec1}
Hyperspectral imaging (HSI) captures detailed spectral information for every pixel in an image, providing a significantly broader spectral range compared to conventional imaging techniques, with channels ranging from tens to hundreds. This unique capability has enabled diverse applications across fields such as science, military, agriculture, and medicine \cite{G2017advances}. However, HSI sensing systems face notable hardware constraints, particularly in miniaturized satellites and resource-limited platforms. These constraints limit the achievable spectral and spatial resolutions. As a result, capturing high-quality, high-resolution hyperspectral images (HR-HSI) is both technically challenging and financially costly, particularly for resource-constrained platforms such as small satellites. Consequently, data scarcity poses a significant hurdle for the practical deployment of many HSI applications.

One widely adopted solution to address these limitations is to capture high-resolution multispectral images (HR-MSIs) alongside low-resolution hyperspectral images (LR-HSIs). Fusion-based super-resolution (SR) techniques \cite{Zhang_2023_CVPR,Ren_2024_CVPR,Chen_2024_CVPR} combine HR-MSIs and LR-HSIs to reconstruct high-resolution hyperspectral images (HR-HSIs) \cite{Zhu2021PZRes,HyperRefiner,HyperTransformer,fusformer,Huang2022DHIF,Min2021MSSJFL,Xiao2021DualUNet}. These techniques have proven effective in overcoming resolution limitations for various HSI-related applications \cite{Dian2021recent, Vivone2023multispectral, Wang2023hyperspectral,rtcs,lee2024prompthsiuniversalhyperspectralimage}.

Despite their effectiveness, HSI fusion tasks present significant challenges that require sophisticated solutions. The primary challenge lies in seamlessly integrating spatial and spectral information while simultaneously preserving fine spectral details and enhancing spatial resolution. Current methods encounter two critical limitations that impede their performance. The first is the progressive loss of essential spatial and spectral information as network depth increases, compromising the quality of the final reconstruction. The second is the insufficient utilization of the rich spatial-spectral relationships inherent in the input data, leaving valuable information unexploited.
These limitations become particularly acute in the context of HSI applications, where the scarcity of high-quality training data makes efficient information utilization paramount. While learning-based HSI fusion methods \cite{Zhu2021PZRes,HyperRefiner,HyperTransformer,fusformer,Huang2022DHIF,Min2021MSSJFL,Xiao2021DualUNet} have significantly advanced the field, even state-of-the-art approaches have yet to fully address fundamental challenges in information loss and data scarcity for HSI-related applications.

To overcome these challenges, we propose HyFusion, a novel dual-coupled framework designed to maximize data utilization and improve HR-HSI reconstruction quality. Our framework employs a Dual-Coupled Network (DCN) architecture that comprises two specialized networks working in tandem: SpeNet for spectral feature extraction and SpaNet for spatial feature extraction. These networks mutually enhance each other during the feature extraction process, enabling effective cross-domain information exchange. At the core of HyFusion is the Enhanced Reception Field Block (ERFB), which integrates Improved Swin Transformer Layers (ISTL) with dense connections. This integration serves two crucial purposes: expanding the receptive field to capture long-range dependencies and preserving critical spatial-spectral features through comprehensive feature map reuse. This design effectively addresses the limitations of existing methods by mitigating information loss and enhancing feature retention. To ensure optimal reconstruction quality, we employ task-specific spectral-aware and frequency-aware loss functions that align reconstructed HR-HSIs with ground-truth data in both frequency and spatial-spectral domains, thereby enhancing fidelity and minimizing artifacts in the final reconstructions.

In summary, our key contributions are as follows:

\begin{itemize} 
    \item We propose HyFusion, a novel framework designed to enhance receptive fields and maximize feature map reuse through the ERFB, which combines shifting-window attention and dense connections to effectively capture long-range dependencies while reducing information loss, thereby boosting data efficiency.
    \item  The proposed DCN dynamically extracts high-frequency spectral and spatial features from LR-HSI and HR-MSI, ensuring efficient cross-domain fusion. Each specialized network leverages complementary information from the other input modality to enhance feature extraction quality.
    \item We design task-specific loss functions to ensure alignment between reconstructed HR-HSIs and ground truth in both frequency and spatial-spectral domains, improving fidelity and eliminating artifacts. 
\end{itemize}


\begin{figure}
    \centering
    \includegraphics[width=0.49\textwidth]{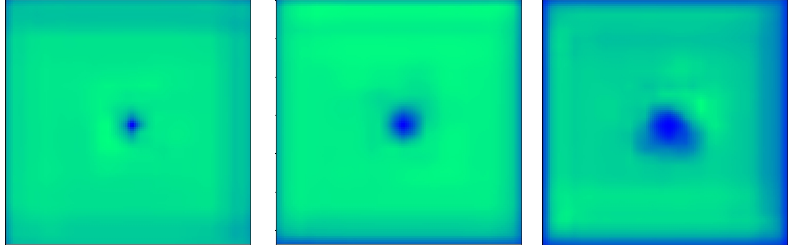}
    \vspace{-0.75cm}   
    \caption{Visualization of Effective Reception Field. (a) naive Swin Transformer Layer (STL), (RSTB in deep feature extraction \cite{SwinIR}); (b) with dense connection (using RDG \cite{Hsu_2024_CVPR}); (c) with dense connection and improved STL (Ours).}\vspace{-0.6cm} 
    \label{fig:receptionfield} 
\end{figure}

\section{Proposed Method}
\label{sec:method}

\begin{figure*}
    \centering
    \includegraphics[width=0.95\textwidth]{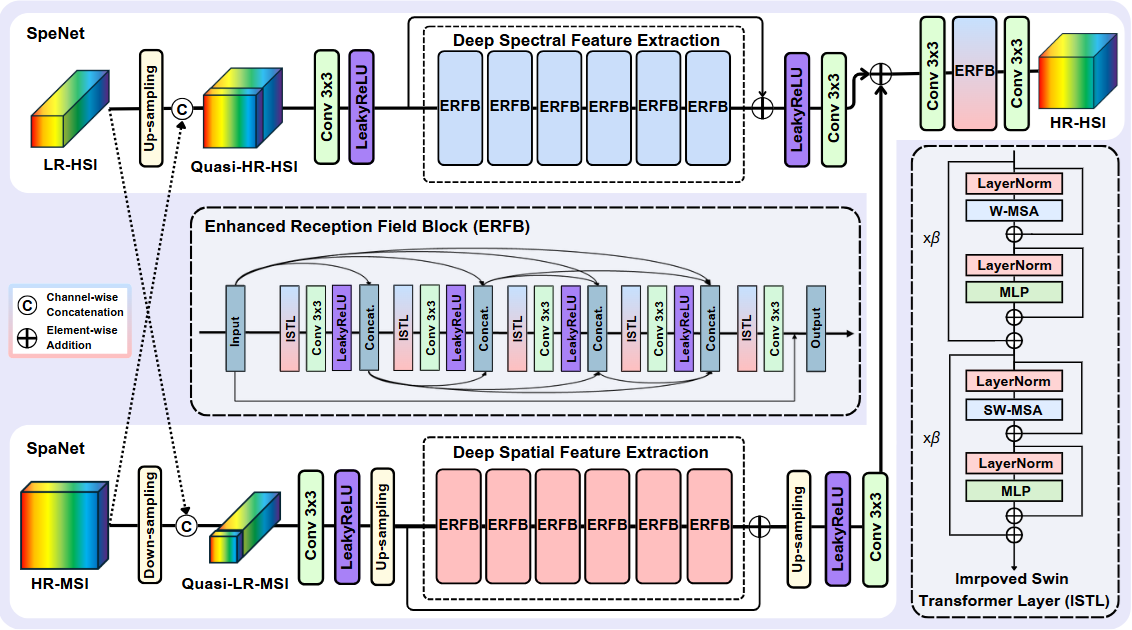}
\vspace{-0.35cm}   
    \caption{The proposed HyFusion network architecture integrates an Enhanced Reception Field Block (ERFB) with an Improved Swin Transformer Layer (ISTL) for HSI/MSI fusion. Through our Dual-Coupled Network (DCN), the architecture learns spatial-spectral representations across two specialized branches. The combination of ERFB, ISTL and DCN enables effective capture of long-range dependencies while maintaining streamlined design, leading to superior fusion results through efficient information exchange between spatial and spectral domains, thereby improving data efficiency for HSI real-world applications.}\vspace{-0.45cm} 
    \label{fig:flowchart} 
\end{figure*}

\subsection{Architecture of HyFusion}

The receptive field plays a crucial role by capturing long-range dependencies and spatial-spectral correlations, as illustrated in Figure \ref{fig:receptionfield}. Given the inherent data scarcity in hyperspectral imaging due to hardware limitations and acquisition costs, maximizing the utility of limited available data becomes crucial. To address this challenge, we propose an efficient LR-HSI/HR-MSI fusion model that leverages dense-connections \cite{huang2018densely} to enable thorough feature reuse throughout the network. This design principle minimizes the loss of valuable spectral and spatial information during fusion while enhancing the reception field, thereby ensuring effective information complementarity and high-quality HR-HSI reconstruction. The feature reuse mechanism is particularly valuable in HSI applications where training data is limited, as it enables the network to make more effective use of each available sample.

Cross-domain information fusion requires effective integration while managing computational and hardware resource demands of large HSI datasets. This necessitates a network architecture that is both simple and efficient. Therefore, Our proposed DCN, which processes LR-HSI and HR-MSI through specialized subnetworks that mutually enhance each other during feature extraction. This coupled design enables the spatial network to leverage spectral information for better pattern recognition, while the spectral network utilizes spatial context to improve feature discrimination. By facilitating this complementary information exchange, DCN enhances representational capacity while maintaining architectural simplicity.

\subsection{Dual-Coupled Network}

This section illustrates the architectural design of our DCN, as shown in Figure \ref{fig:flowchart}, including the spatial and spectral perception networks (SpaNet and SpeNet) for LR-HSI and HR-MSI. To effectively combine the features of LR-HSI and HR-MSI, the proposed SpaNet and SpeNet not only extract their corresponding feature representations but also utilize early image fusion to allow the corresponding counterparts to provide guidance simultaneously. This approach allows us to seamlessly integrate spatial and spectral feature representations without increasing model complexity.

Let the LR-HSI and HR-MSI be $\mathbf{X}_{h} \in \mathbb{R}^{h_h\times w_h\times b}$ and $\mathbf{X}_m \in \mathbb{R}^{h\times w\times b_m}$. The observable LR-HSI can be modeled as $\mathbf{X}_h = \mathbf{Y}\mathbf{B}$, where $\mathbf{B}$ is a blurring matrix that reduces the pixel count. Similarly, the observable HR-MSI is given by $\mathbf{X}_m = \mathbf{D}\mathbf{Y}$, where $\mathbf{D}$ is a downsampling matrix that decreases the number of spectral bands.

The reconstructed HR-HSI denotes $\mathbf{Y}^* \in \mathbb{R}^{w\times h\times b}$ by:
\begin{gather}
    {\mathbf{Z}_{hm}}=f_{\text{DCN}}(\mathbf{X}_h, \mathbf{X}_m)= \text{Concat}.(\mathbf{Z}_{h}, \mathbf{Z}_{m}),
\end{gather}
where the high-spectral-resolution and high-spatial-resolution features are obtained as:
\begin{equation}
    \mathbf{Z}_{h} = f_{\text{SpeNet}}(\mathbf{X}_h,\mathbf{X}_m);\\
    \mathbf{Z}_{m} = f_{\text{SpaNet}}(\mathbf{X}_m,\mathbf{X}_h)
\end{equation}

For SpeNet, we up-sample the LR-HSI to obtain $\mathbf{X}_{h}^{u} = f_{\text{up}}(\mathbf{X}_{h})$, where $f_{\text{up}}$ denotes the bilinear interpolation function. Then, the quasi-spectral draft is obtained by:
\vspace{-0.2cm} 
\begin{equation}
    \mathbf{Z}'_{h} = \text{Concat}.(\mathbf{X}_{h}^{u}, \mathbf{X}_m)
    \vspace{-0.2cm} 
\end{equation}

Similarly for SpaNet, we down-sample the HR-MSI to obtain $\mathbf{X}_{m}^{d} = f_{\text{down}}(\mathbf{X}_{m})$, where $f_{\text{down}}$ denotes the bicubic interpolation function, creating a quasi-spatial draft:
\vspace{-0.2cm} 
\begin{equation}
    \mathbf{Z}_{m}^{'} = \text{Concat}.(\mathbf{X}_{h}, \mathbf{X}_m^{d})\vspace{-0.2cm} 
\end{equation}

After obtaining these preliminary drafts, shallow feature extraction is performed using $3 \times 3$ convolution with LeakyReLU activation (negative slope $0.2$). The ERFB is then used for deep feature extraction. The reconstructed HR-HSI is:
\vspace{-0.2cm} 
\begin{equation}
    {\mathbf{Y}^*}=f_{\text{Rec}}({\mathbf{Z}_{hm}}) \vspace{-0.2cm} 
\end{equation}

\begin{table*}[ht]
\centering
\caption{Performance evaluation and complexity comparison of the proposed method and other fusion models. The best results are highlighted in \best{bold-red}, while the second-best results are highlighted in \secondBest{bold-blue}. Rows are colored to distinguish different approaches: \colorbox{cnn_color}{CNN-based}, \colorbox{transformer_color}{Transformer-based}, and \colorbox{ours_color}{the proposed} methods.}
\label{tab:performance}\vspace{-0.25cm} 
\scalebox{0.83}{
\begin{tabular}{r|cccc|cccc|cccc}  
\toprule[0.15em]
\multirow{2}{*}{\textbf{Method}} &  \multicolumn{4}{c|}{Complexity} & \multicolumn{4}{c|}{4 Bands LR-HSI} & \multicolumn{4}{c}{6 Bands LR-HSI}\\
 &  \#Params↓ & FLOPs↓ & Run-time↓ & Memory↓ & PSNR↑ &  SAM↓ &  RMSE↓  & ERGAS↓  &  PSNR↑ &  SAM↓ &  RMSE↓ & ERGAS↓   \\\hline
\rowcolor{cnn_color}
\textbf{PZRes-Net} \cite{Zhu2021PZRes} {\tt\small{TIP'21}} & 40.15M & 5262G & 0.0141s & 11059MB & 34.963 &  1.934  & 35.498 & 1.935& 37.427 &  1.478 &  28.234 &  1.538  \\
\rowcolor{cnn_color}
\textbf{MSSJFL} \cite{Min2021MSSJFL} {\tt\small{HPCC'21}} & 16.33M & 175.56G & \secondBest{0.0128s} & \best{1349MB} &  34.966 & 1.792 &  33.636 &  2.245 &  38.006 &  1.390 &  26.893 & 1.535 \\
\rowcolor{cnn_color}
\textbf{Dual-UNet} \cite{Xiao2021DualUNet} {\tt\small{TGRS'21}} & \secondBest{2.97M} & \secondBest{88.65G} & \best{0.0127s} & \secondBest{2152MB} & \secondBest{35.423} & 1.892 & 33.183 & \secondBest{1.796} & 38.453 & 1.548 & 26.148 & 1.205 \\
\rowcolor{cnn_color}
\textbf{DHIF-Net} \cite{Huang2022DHIF} {\tt\small{TCI'22}} & 57.04M & 13795.11G &  6.005s & 29381MB &34.458 & 1.829 & 34.769 & 2.613 & \secondBest{39.146} & 1.239 & 25.309 & \secondBest{1.113} \\
\rowcolor{transformer_color}
\textbf{FusFormer} \cite{fusformer} {\tt\small{TGRS'22}} & \best{0.18M} & \best{11.74G} & 0.0158s & 5964MB & 34.217 & 2.012 & 35.687 & 1.996 & 38.637 & 1.678 & 28.674 & 1.204 \\
\rowcolor{transformer_color}
\textbf{HyperTransformer} \cite{HyperTransformer} {\tt\small{CVPR'22}} & 142.83M & 343.96G & 0.0252s & 8104MB & 28.692 &  3.664 &  62.231 & 4.774 & 32.954 & 2.568 & 41.256 & 3.834 \\
\rowcolor{transformer_color}
\textbf{HyperRefiner} \cite{HyperRefiner} {\tt\small{TJDE'23}} & 19.32M & 94.37G & 0.0237s & 7542MB & 33.298 & 2.129 & 38.769 & 2.086 & 37.654 & 1.590 & 29.629 & 1.403  \\
\rowcolor{cnn_color}
\textbf{QRCODE} \cite{QRCODE} {\tt\small{TGRS'24}} & 41.88M & 2231.19G & 0.2452s & 15028MB & 35.361 & \secondBest{1.623} &  \secondBest{32.711} & 2.027 & 38.948 & \secondBest{1.148} &  \secondBest{24.617} & 1.429 \\ \hline
\rowcolor{ours_color}
\textbf{HyFusion (Ours)}& 25.53M & 641.49G &  0.0294s & 2572MB & \best{37.421} &  \best{1.256} &  \best{27.425} &  \best{1.172}& \best{40.197} & \best{1.017} &  \best{21.455} &  \best{1.076}  \\
\bottomrule[0.15em]
\end{tabular}}\vspace{-0.55cm} 
\end{table*}

\subsection{Enhanced Reception Field Block}
Given the inherent data scarcity in hyperspectral imaging due to hardware limitations and acquisition costs, maximizing the utility of limited available data becomes crucial. The proposed ERFB is specifically designed to address this challenge along with two fundamental issues in hyperspectral image fusion: capturing comprehensive spatial-spectral relationships through long-range dependencies and maximizing feature utilization efficiency. By effectively reusing features and enhancing the receptive field, ERFB enables the network to learn robust representations even with limited training data.

Our design philosophy focus on a dual-enhancement strategy that synergistically combines dense connections for thorough feature reuse with an ISTL for enhanced receptive field and long-range dependency modeling. The dense connection structure ensures that each layer can access and build upon all previous feature representations, effectively preventing information loss as the network depth increases. This is particularly crucial for hyperspectral image fusion, where preserving fine spectral details while maximizing the utility of limited data is essential.

At the core of our ERFB is the ISTL, which consists of two sequential processing paths: W-MSA and SW-MSA. Each path contains LayerNorm, attention (W-MSA or SW-MSA), LayerNorm, and MLP blocks with learnable parameters $\beta_1$ and $\beta_2$ respectively to adaptively control feature contributions:

\vspace{-0.1cm} 
\begin{equation}
\text{ISTL}(\mathbf{X}) = \beta_1 \cdot \text{Path}_\text{W-MSA}(\mathbf{X}) + \beta_2 \cdot \text{Path}_\text{SW-MSA}(\mathbf{X})
\vspace{-0.1cm} 
\end{equation}

The dense connection pattern in ERFB progressively accumulates and reuses features for reception field enhancement:
\vspace{-0.1cm} 
\begin{equation}
\mathbf{Z}_j = \text{Cat}(\text{Conv}_{3\times3}(\text{ISTL}(\mathbf{Z}_0, ..., \mathbf{Z}_{j-1}))), j = 1, 2, 3, 4
\vspace{-0.1cm} 
\end{equation}

\noindent where $\text{Cat}(\cdot)$ represents the channel-wise concatenation, and $\text{Conv}_{3\times3}$ fuses the concatenated features with LeakyReLU activation with $0.2$ negative slope. The final stage processes the accumulated features:

\vspace{-0.1cm} 
\begin{equation}
\mathbf{Z}_5 = 0.2 \times \text{Conv}_{3\times3}(\text{ISTL}(\mathbf{Z}_4)) + \mathbf{Z}_{0}
\vspace{-0.1cm} 
\end{equation}

With the efficient information flow via dense connectivity within ERFB, and adaptive feature weighting through ISTL, the long-range dependency modeling capability of the proposed HyFusion can be improved, as shown in Figure \ref{fig:receptionfield}. In the context of limited hyperspectral training data, this architecture proves particularly valuable as it maximizes the utility of available information through feature reuse while maintaining the ability to capture complex spatial-spectral relationships. When incorporated into both SpeNet and SpaNet branches of the DCN architecture, this approach provides a informative representations for high-quality hyperspectral image fusion while maintaining computational and data efficiency.

\subsection{Task-Specific Optimization}

Recognizing the paramount importance of spectral index in HSI, we introduce a Spectral Angle Mapper (SAM) loss function and Stationary Wavelet Transform (SWT) loss to guide the network:

\vspace{-0.1cm} 
\begin{equation}
    \ell_{\text{SAM}} = \frac{1}{N} \sum_{n=1}^{N} \cos^{-1} \left( \frac{\mathbf{Y}_n^T \mathbf{Y}^*_n }{\|\mathbf{Y}_n\|_2 \cdot \|\mathbf{Y}_n^*\|_2} \right)
    \vspace{-0.1cm} 
\end{equation}

\vspace{-0.1cm} 
\begin{equation}
\ell_{\text{SWT}}(\mathbf{Y}, \mathbf{Y}^*) = \mathbb{E} \bigl[ \sum_{j} \lambda_{j} \big\| \text{SWT}(\mathbf{Y})_{j} - \text{SWT}(\mathbf{Y}^*)_{j} \big\| _1  \bigr]
\vspace{-0.1cm} 
\end{equation}

Finally, the total loss with L1 loss can be written as:
\vspace{-0.1cm} 
\begin{equation}
    \ell_{\text{Total}}(\mathbf{Y}, \mathbf{Y}^*) = \ell_{\text{1}}(\mathbf{Y},\mathbf{Y}^*) + \lambda_{1} \ell_{\text{SAM}}(\mathbf{Y}, \mathbf{Y}^*)+ \lambda_{2} \ell_{\text{SWT}}(\mathbf{Y}, \mathbf{Y}^*)
    \vspace{-0.1cm} 
\end{equation}

\section{Experimental Results}
\label{sec:expALL}

\subsection{Experiment Settings}

\subsubsection{Dataset Illustration}
The dataset used for performance evaluation in this study was acquired by the Airborne Visible/Infrared Imaging Spectrometer (AVIRIS) sensor \cite{Vane1993airborne}. It consisted of 2,078 HR-HSI images, which were randomly partitioned into training, validation, and testing sets for performance evaluation. The training set contained 1,678 images, while the validation and testing sets contained 200 images for each. The spatial and spectral resolutions of the HR-MSI and LR-HSI were $256\times 256 \times M_m$ and $64\times 64\times 172$, respectively, where $M_m$ is either 4 or 6 in our experiments.

\subsubsection{Implementation Details} The proposed method was implemented using the PyTorch deep learning framework. The batch size was set to 4, and the number of training epochs was fixed to 600 for all experiments involving the proposed method. For the peer methods, the number of training epochs was set according to their default values as specified in their respectively original publications. The ADAM \cite{Kingma2014adam} optimizer was used for training, with an initial learning rate of 0.0001. The learning rate was adjusted during the training process using the Cosine Annealing learning decay scheduler. 
The weights of the penalty terms in the loss function, denoted as $\lambda_{1}$ and $\lambda_{2}$, were both set to 0.01. Standard data augmentation, including random cropping and rotation, is adopted in this paper for all of the evaluated methods.

\begin{figure}
    \centering
    \includegraphics[width=0.49\textwidth]{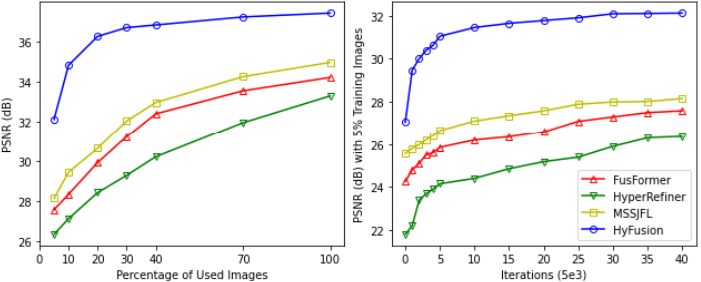}
    \vspace{-0.75cm}   
    \caption{Performance comparison between several HSI fusion models. (a) Evaluation of model performance with varying training data sizes. (b) Validation performance curves during training with 5\% training samples.}\vspace{-0.6cm} 
    \label{fig:de} 
\end{figure}

\subsection{Quantitative Results}

To evaluate the performance of our hyperspectral image fusion method, we compared it with five other supervised hyperspectral image fusion methods: PZRes-Net \cite{Zhu2021PZRes}, MSSJFL \cite{Min2021MSSJFL}, Dual-UNet \cite{Xiao2021DualUNet} and DHIF-Net \cite{Huang2022DHIF}, FusFormer \cite{fusformer}, HyperTransformer \cite{HyperTransformer}, HyperRefiner \cite{HyperRefiner}, QRCODE \cite{QRCODE}. The performance was objectively measured using PSNR, SAM, and RMSE, ERGAS. Experiments were conducted with both 4 and 6 MSI bands. Table \ref{tab:performance} present the results of comparing our proposed method with the state-of-the-art methods.

The experimental results demonstrate that the proposed framework outperforms the compared state-of-the-art methods in terms of spectral reconstruction performance.

\subsection{Data Efficiency Analysis}

To evaluate HyFusion's ability to learn from limited training data, we conducted experiments comparing our approach with state-of-the-art methods FusFormer \cite{fusformer}, HyperRefiner \cite{HyperRefiner} and MSSJFL \cite{Min2021MSSJFL} using 4 bands LR-HSI for fusion. As shown in \ref{fig:de}(a), we gradually reduced the percentage of training samples from 100\% to 5\%. HyFusion outperforms other methods by a significant margin. This demonstrates our model's strong data efficiency, enabled by the feature reuse through dense connections and enhanced learning capability from ISTL.

The convergence analysis in \ref{fig:de}(b) further validates HyFusion's advantages. Our model not only achieves faster convergence but also reaches higher PSNR values throughout the training process. This rapid convergence can be attributed to the effective spatial-spectral feature extraction through our dual-branch architecture and the improved gradient flow from dense connections. The results highlight that HyFusion's architectural innovations lead to both better performance and more efficient learning from limited hyperspectral training data.

\section{Conclusion}
\label{sec:conclusion}

In this paper, we have proposed HyFusion, a novel framework that effectively addresses fundamental challenges in hyperspectral image fusion through its dual-coupled network architecture. By designing specialized subnetworks that mutually enhance each other during feature extraction, HyFusion enables efficient cross-domain information exchange between HR-MSI and LR-HSI inputs. The Enhanced Reception Field Block (ERFB) serves as a crucial component, integrating dense connections with an Improved Swin Transformer Layer (ISTL) to maximize feature utilization and expand the receptive field. 
Our experiments demonstrate that HyFusion achieves state-of-the-art performance across multiple evaluation metrics while maintaining competitive computational efficiency. Particularly noteworthy is the framework's strong performance with limited training data, showcasing its effectiveness in addressing the inherent data scarcity challenges in hyperspectral imaging.

\small
\bibliographystyle{IEEEtranN}
\bibliography{sn-bibliography}	
\end{document}